\documentclass[prb,twocolumn,showpacs,amssymb]{revtex4}

\usepackage{graphicx}
\usepackage{dcolumn}
\usepackage{bm}

\begin{document}


\title{Atoms in the anionic domain, $Z < N$}
\author{Gabriel Gil and Augusto Gonzalez}
\affiliation{Instituto de Cibernetica, Matematica
 y Fisica, Calle E 309, Vedado, Ciudad Habana, Cuba}

\begin{abstract}
We study atoms with $N$ electrons, and nuclear charge $Z$. It is well known that the
cationic regime, $Z > N$ is qualitatively described by Thomas-Fermi theory. The anionic
regime, $Z < N$, on the other hand, is characterized by an instability threshold at 
$Z_c\lesssim N-1$,
below which the atom spontaneously emits an electron. We compute the slope of the energy 
curve at $Z=N-1$ by means of a simple model that depends on the electron affinity and the
covalent radius of the neutral atom with $N-1$ electrons. This slope is used in order to
estimate $Z_c$, which is compared with previous numerical results. Extrapolation of the
linear behaviour in the opposite direction, up to $Z=N$, allows us to estimate the 
ionization potential of the atom with $N$ electrons. The fact that the obtained 
ionization potentials are qualitatively correct is an indication that, with regard to certain 
properties, neutral atoms are closer to the anionic instability threshold than they 
are to the Thomas-Fermi, large $Z$, regime. A regularized series is written for the ionization
potential that fits both, the large $Z$ and $Z\to Z_c$ regimes.
\end{abstract}


\maketitle

\section{Introduction}
\label{Introduction}

Qualitative understanding of phenomena come from very simple models that capture their essence. In atoms, perhaps the most successful simple models are the famous Thomas-Fermi (TF) theory \cite{TF1,TF2}, and the large-$D$ expansion \cite{largeD}, leading, the latter, to a qualitative picture similar to the one devised by Lewis long ago \cite{Lewis}.

The theory by Thomas and Fermi is a mean-field one in which the scaling properties of the kinetic energy and Coulomb potentials are apparent. This fact leads to a universal relation for the energy of the atom with $N$ electrons and charge $Z$, \cite{TF2} $E_{TF}(N,Z)=N^{7/3}f(N/Z)$, or for the ionization potential \cite{ourPRA}, $I_p(N,Z)=N^{4/3}g(N/Z)$, where the universal functions $f$ and $g$ depend only on the combination $N/Z$.

\begin{figure}[t]
\begin{center}
\includegraphics[width=.9\linewidth,angle=0]{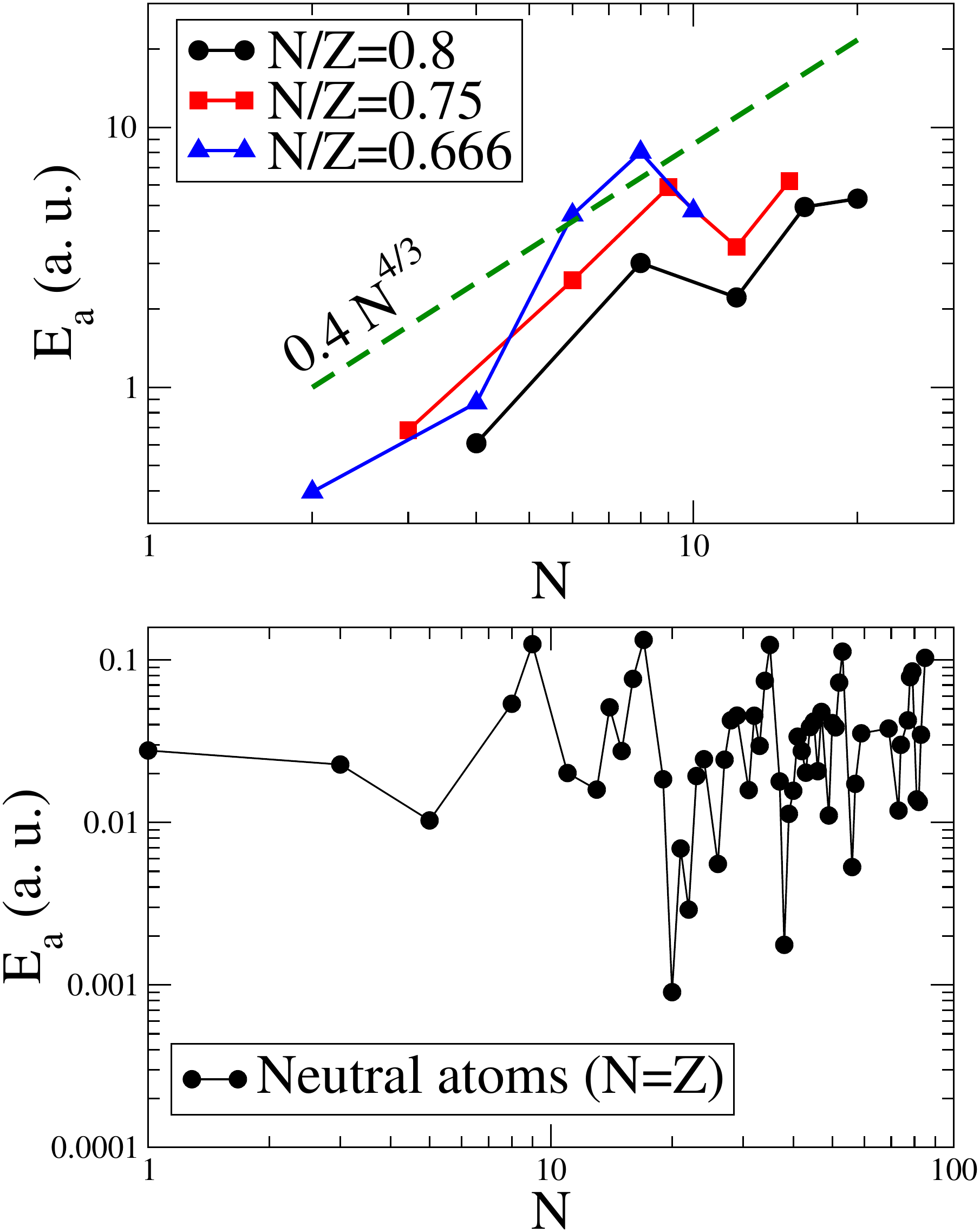}
\caption{\label{fig1} (Color online) Upper panel: Electron affinities of cations with $N\le 20$ and fixed $N/Z$ ratios, obtained from the ionization potential of the cation with $N+1$ electrons. Data are taken from the NIST database \cite{NIST}. The line $0.4 \sim N^{4/3}$ is drawn as a reference for the slope. Lower panel: Electron affinities of neutral atoms \cite{RSC}.}
\end{center}
\end{figure}

According to TF theory, for a fixed ratio $N/Z$, the ionization potential (or the electron affinity, 
$E_a$) should behave as $E_a \sim N^{4/3}$, when $N$ increases. This is the correct dependence for cations ($N<Z$), even for small $N$ values, as can be seen in the upper panel of Fig. \ref{fig1}, but it breaks down for neutral atoms ($N=Z$), where $I_p$ or $E_a$, apart from fluctuations due to shell effects, take roughly a constant value (Fig. \ref{fig1}, lower panel).

The reason for such a breaking down of TF predictions for neutral atoms, in certain situations, is the proximity of the anionic instability threshold, that is, the fact that atoms become unstable when $Z<Z_c\lesssim N-1$.

In our paper, we aim at studying the vicinity of the anionic threshold. We show that a simple physical picture holds at $Z=N-1$, which is close enough to $Z_c$, allowing us to accurately compute the slope of the energy curve. Extrapolating this line, we obtain a quite good estimation of $Z_c$.

The physics of neutral atoms, takes place in an intermediate regime between the TF-region ($Z\gtrsim N$), and the anionic instability region ($Z\lesssim N-1$). Extrapolating the linear dependence up to $Z=N$, we get an estimation of the ionization potential of the atom with $N$ electrons. The fact that this estimation is qualitatively good, could be understood as an indication that, with regard to certain properties, neutral atoms are closer to the anionic instability region than they are to the cationic, Thomas-Fermi, regime.

\section{The $E_b$ vs $Z$ curve}
\label{EbvsZ}

\begin{figure}[t]
\begin{center}
\includegraphics[width=.95\linewidth,angle=0]{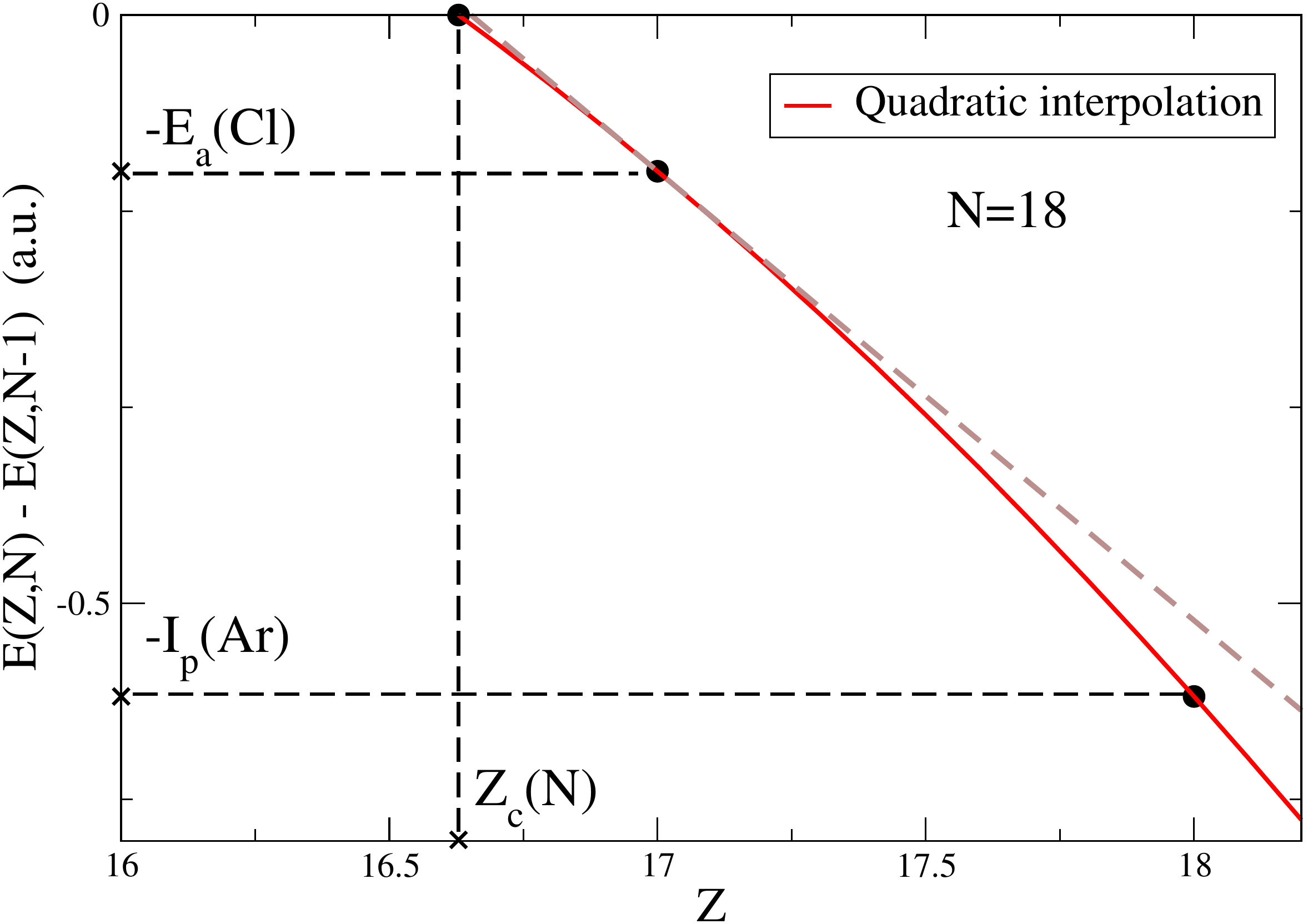}
\caption{\label{fig2} (Color online) $E_b$ vs $Z$ for Ar-like ions. The solid line quadratically interpolates between the experimental points. The dashed line is the tangent curve at $Z=N-1$.}
\end{center}
\end{figure}

We show in Fig. \ref{fig2} the binding energy (with negative sign) of the Ar-like atom ($N=18$) as a function of $Z$, $E_b=E(Z,N)-E(Z,N-1)$. $E(Z,N)$ is the energy of the $N$-electron atom with nuclear charge $Z$. Notice that, when $Z=N$, $E_b$ equals minus the ionization potential of the $N$-electron atom, whereas at $Z=N-1$, $E_b$ is equal to minus the electron affinity of the atom with $N-1$ electrons. 

In the figure, $I_p$(Ar) and $E_a$(Cl) are the experimental values. $Z_c=16.629$ is the estimation by Kais and co-workers \cite{Kais}. The solid line is an interpolation by means of a second-order polynomial. The dashed line is the tangent curve at $Z=N-1$. Notice that this line intersects the $E_b=0$ axis practically at $Z_c$.

In the next sections, we show that there is a way to accurately compute the slope at $Z=N-1$, thus allowing a good estimation of $Z_c$.

On the other hand, the tangent line provides also a crude estimation of the ionization potential of the neutral atom with $N$ electrons, which nevertheless shows the correct dependence on $N$, as it will be demonstrated below. 

\section{The model}
\label{model}  

As mentioned above, at $Z=Z_c$ the atom spontaneously ionizes by emitting an alectron. Thus, near $Z_c$ there should be an electronic orbit with a relatively high characteristic radius. If we think about the Helium atom, for example, we may write the spatial wave function (up to a normalization factor) as:

\begin{equation}
\psi(1,2)=\phi_c(1)\phi_e(2)+\phi_c(2)\phi_e(1),
\label{eq1}
\end{equation}

\noindent
where 1 and 2 refer to electron coordinates, $\phi_c$ is the $s$ orbital of the core electron, and $\phi_e$ - the $s$ orbital of the external electron.

The overlap function between external and core orbitals, $\langle\phi_c|\phi_e\rangle$, should be small. Neglecting overlapping, we may write an effective Schrodinger equation for $\chi=\phi_e r$ which, in the general case, reads:

\begin{equation}
\left\{-\frac{1}{2}\frac{{\rm d}^2}{{\rm d}r^2}+V(r)+\Theta(r-R)\frac{\alpha}{r}+\frac{l(l+1)}{2 r^2}\right\}\chi=E_b \chi.
\label{eq2}
\end{equation}

\noindent
$V(r)$ takes account of the short-range interactions with the core electrons. It's effective radius, $R$, should be similar to the radius of the $N-1$ electron atom. The long-range Coulomb interaction, acting only for $r>R$, $\Theta$ is the step function, exhibits an effective charge $-\alpha=Z-(N-1)$. $l$ is the angular momentum of the external orbital.

On rigorous terms, it is known \cite{lineal} that $N-2<Z_c<N-1$ and that the dependence of $E_b$ on $Z$ is linear near $Z_c$. The later stament is equivalent to saying that $\phi_e$ remains normalized at threshold ($\chi\sim \exp-2\sqrt{\alpha_c r}$), thus, by the Feynman-Hellman theorem, the slope can be computed as $-\langle \phi_e|\Theta(r-R)1/r|\phi_e\rangle/\langle \phi_e|\phi_e\rangle$. The case $Z_c=N-1$, that is $E_a(N-1)=0$, is special.

We shall compute the slope of the enery curve not at $Z_c$, but when $\alpha=0$ and the core is strictly neutral. In that case, $V$ supports a bound state at $-E_a$, and $\chi$ at large distances behave as:

\begin{equation}
\chi\sim \exp(-\kappa r), 
\label{eq3}
\end{equation}

\noindent
where $\kappa=\sqrt{2 E_a}$. The term depending on $l$ in Eq. (\ref{eq2}) gives rise to corrections that decay still faster with $r$.

We will use Eq. (\ref{eq3}) in order to compute the slope. It corresponds to a kind of zero-range forces theory \cite{ZRF}, which, in principle, should be valid when the effective radius of $\chi$, i.e. $1/\sqrt{2 E_a}$, is greater than $R$. The details of the wave function for $r<R$ are not important. We will show below how good this approximation is. For the slope, we get:

\begin{equation}
s=\frac{\int_R^{\infty}dr~e^{-2\kappa r}/r}{\int_R^{\infty}dr~e^{-2\kappa r}}=-2\kappa F(2\kappa R), 
\label{eq4}
\end{equation}

\noindent
where $F(x)=e^x\int_x^{\infty}dy~e^{-y}/y$. Turning back to the Ar example, shown in Fig. \ref{fig2}, we notice that the slope that follows from the quadratic interpolation is -0.3818. If we use the experimental $E_a$(Cl)=0.1328 a.u., and the covalent radius of Cl, $R_{cov}$(Cl)=1.8897 a.u., as an estimation of $R$, we get from Eq. (\ref{eq4}) a slope of -0.3800, quite close to the actual one. 

\begin{figure}[t]
\begin{center}
\includegraphics[width=.95\linewidth,angle=0]{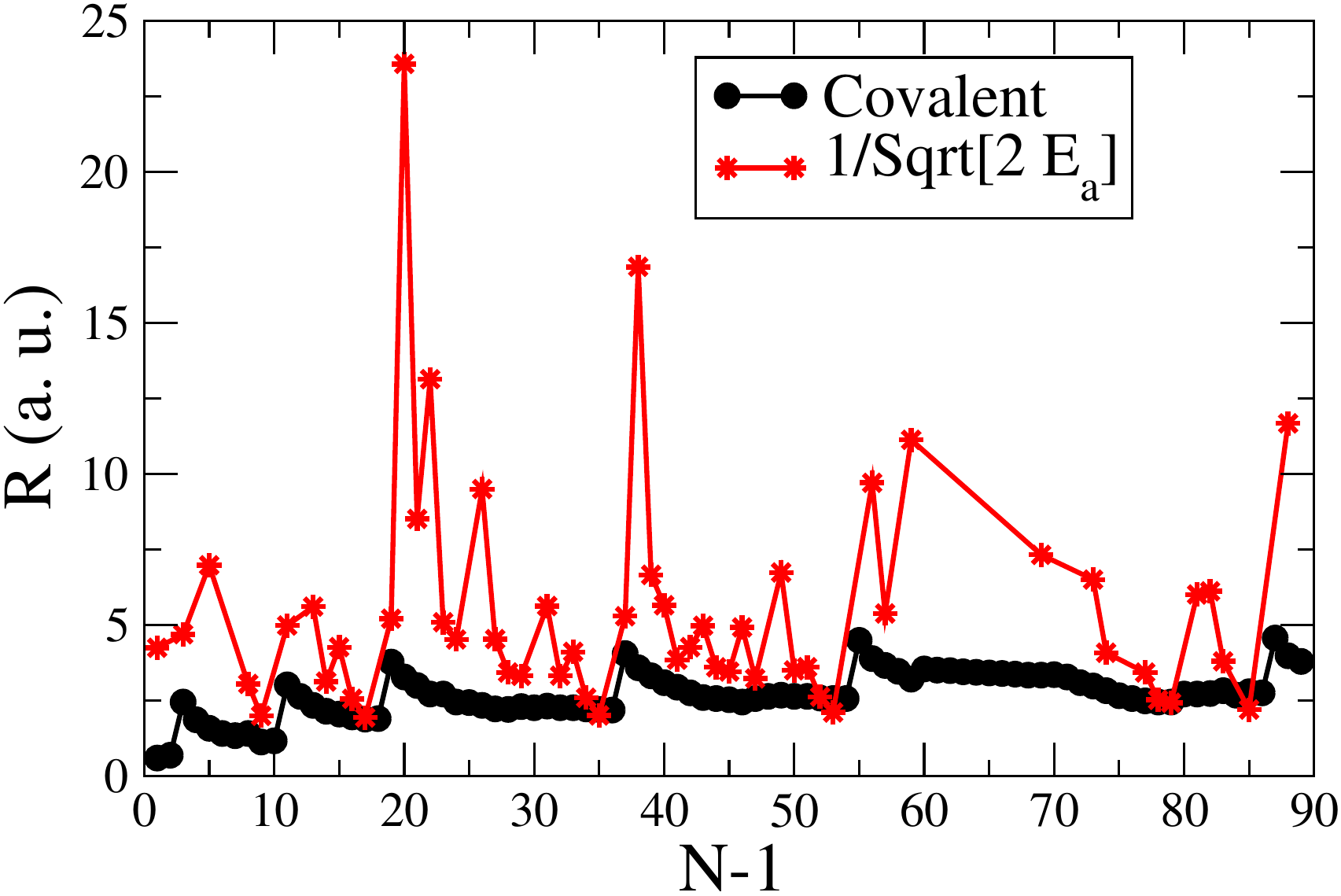}
\caption{\label{fig3} (Color online) Comparison between the effective radii of the external orbitals, $1/\sqrt{2 E_a}$, and the covalent radii of the neutral atoms with $N-1$ electrons.}
\end{center}
\end{figure}

In Fig. \ref{fig3}, we take the covalent radii of atoms with $N-1$ electrons \cite{RSC} as a measure of the core sizes, and compare them with the effective radii of the external orbitals, given by $1/\sqrt{2 E_a}$. This figure shows that the zero-range forces theory is actually a good approximation in the present case.

\section{Threshold nuclear charges}

According to Eq. (\ref{eq4}), in the neighborhood of $\alpha=0$, we may write:

\begin{equation}
E_b\approx-E_a+2 \alpha\kappa F(2\kappa R).
\label{eq5}
\end{equation}

The threshold nuclear charge, $Z_c$, is the charge at which $E_b=0$. If we write $Z_c=N-1-g_c$, then for $g_c$ one gets:

\begin{equation}
g_c\approx\sqrt{E_a}/(2\sqrt{2} F(2\kappa R)).
\label{eq6}
\end{equation}

We compare the estimation given by Eq. (\ref{eq6}) with numbers coming from an interpolation scheme by Kais and co-workers \cite{Kais} (see Appendix). The latter is consistent with numerical computations for atoms with up to 18 electrons \cite{Hogreve}. In quality of $R$, we use the covalent radii, as before.

\begin{figure}[t]
\begin{center}
\includegraphics[width=.95\linewidth,angle=0]{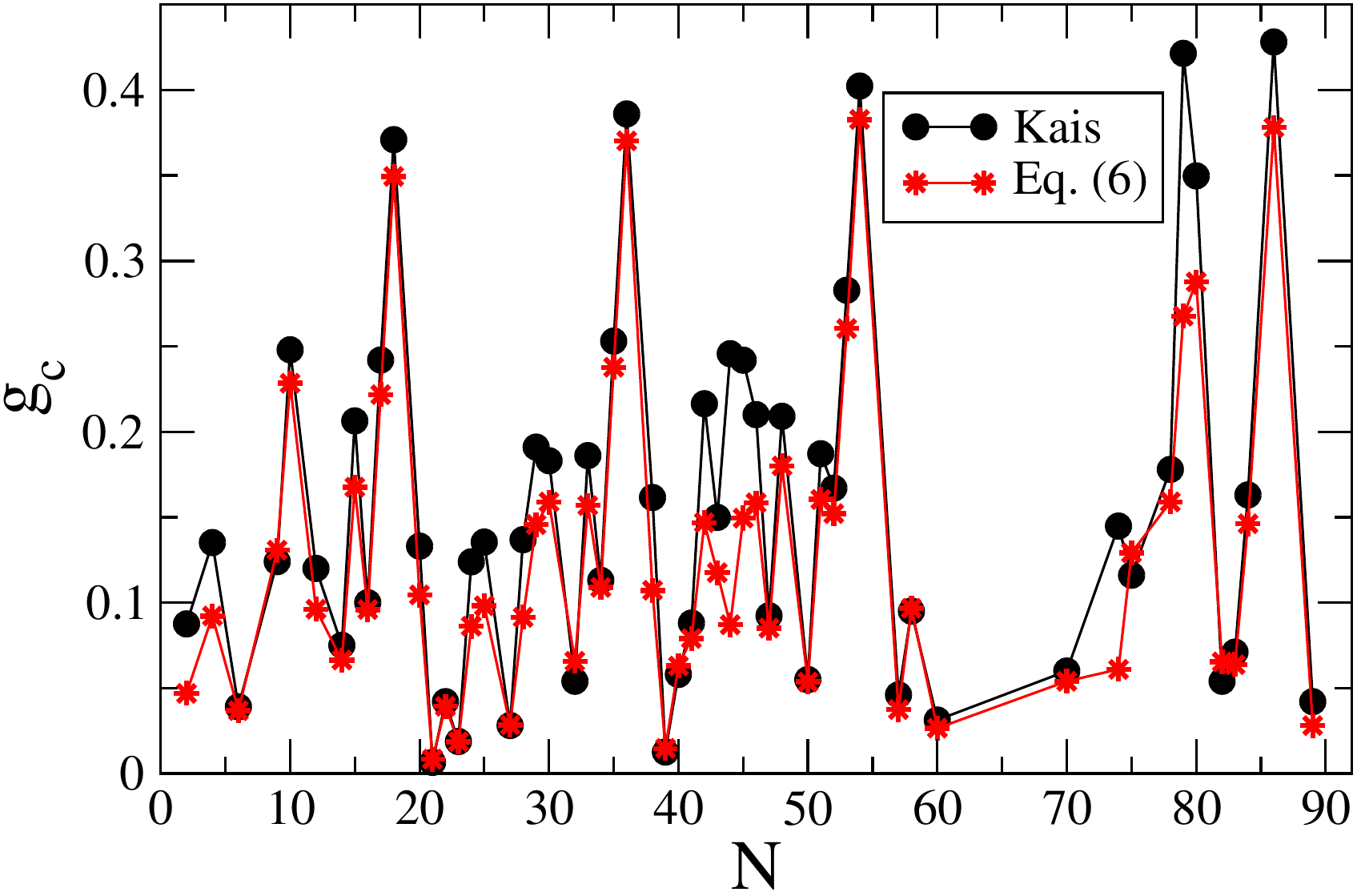}
\caption{\label{fig4} (Color online) Computed values of $g_c$, according to the Kais scheme \cite{Kais} (see Appendix), versus the result coming from Eq. (\ref{eq6}).}
\end{center}
\end{figure}

The results are displayed in Fig. \ref{fig4}. The agreement is quite good, in spite of the simplicity of Eq. (\ref{eq6}).
Certain remarkable deviations as, for example, the case of W ($N=74$), could be related to the oversimplification 
leading to our Eq. (\ref{eq6}), or deficiencies in the Kais interpolation procedure. 

\section{The ionization potential of neutral atoms}

We may use Eq. (\ref{eq5}) with $Z=N$, that is $\alpha=-1$, in order to obtain a crude estimation of the ionization potential:

\begin{equation}
I_p=E_a+2\kappa F(2\kappa R).
\label{eq7}
\end{equation}

\begin{figure}[t]
\begin{center}
\includegraphics[width=.95\linewidth,angle=0]{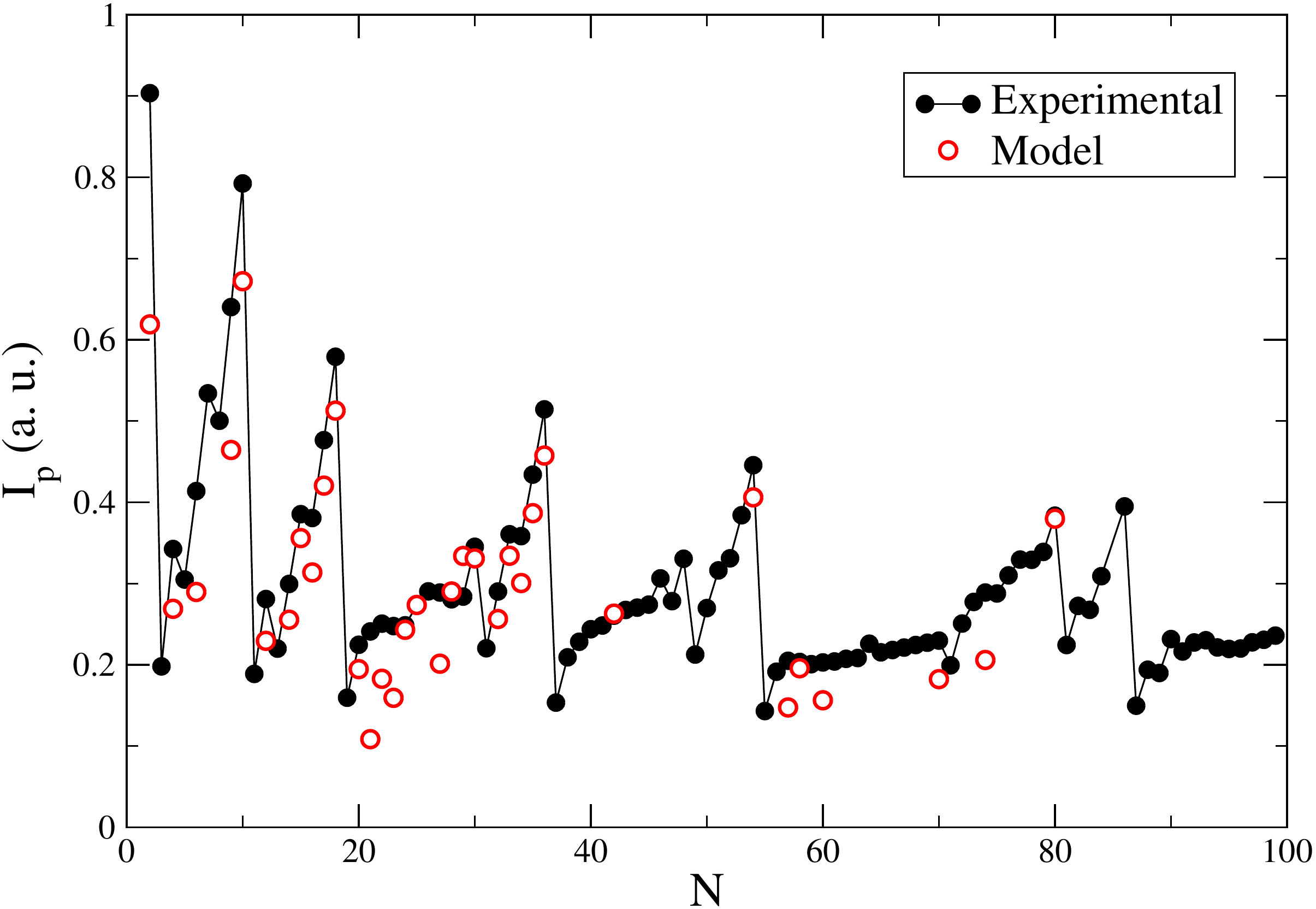}
\caption{\label{fig5} (Color online) Comparison between experimental ionization potentials \cite{NIST} and the estimation coming from Eq. (\ref{eq7}).}
\end{center}
\end{figure}

\noindent
Comparison with the experimental data is made in Fig. \ref{fig5}. It is apparent that the qualitative dependence of $I_p$ on $N$ is reproduced by this simple expression, although deviations are noticeable.

The fact that $I_p$ can be qualitatively reproduced from the linear dependence of $E_b$ on $Z$, coming from the anionic instability threshold, can be understood as an indication that, at least with regard to this property, the physical regime under which neutral atoms operate is closer to the instability threshold than to the large-$Z$ regime, described by TF theory.

A more accurate and general expression for $E_b$, which takes account of the quadratic dependences on $Z$, coming from the large-Z regime, is given in the next section.

\section{Renormalized perturbative series for $E_b$}

The large-$Z$ series for the magnitude $E_b$ has the form:

\begin{equation}
E_b=-a_2 Z^2+a_1 Z+\cdots,
\label{eq8}
\end{equation}

\noindent
where the coeffcients $a_2$ and $a_1$ can be easily computed. Indeed, $a_2$ comes from the energy of non-interacting electrons in the nuclear field. It is exactly $1/(2n_f^2)$, where $n_f$ is the principal quantum number of the last occupied shell. On the other hand, $a_1$ comes from the perturbative evaluation of Coulomb repulsion between electrons:

\begin{equation}
a_1=\sum_{j\ne i}\{ \langle i,j||i,j\rangle-\langle i,j||j,i\rangle\},
\end{equation}

\noindent
where the sum runs over occupied orbitals, $j$, and $i$ is the orbital left by the electron. $\langle i,j||i,j\rangle$ and $\langle i,j||j,i\rangle$ are, respectively, direct and exchange two-electron integrals between hydrogen orbitals $i$ and $j$.

For large $N$, we obtain\cite{largeZ}: $a_2\approx 1/2~(2/3)^{2/3}N^{-2/3}$, and $a_1\approx 0.72~N^{1/3}$, according to the scaling predicted by TF theory \cite{ourPRA}.

Following an idea used in other contexts \cite{renormS}, we shall regularize the series given in Eq. (\ref{eq8}) by introducing the next terms of the expansion:

\begin{equation}
E_b=-a_2 Z^2+a_1 Z+a_0+a_{-1}/Z,
\label{eq10}
\end{equation}

\noindent
and requiring that the regularized series fulfill certain conditions in the anionic domain. In particular, at $Z=N-1$, we ask for $E_b=-E_a$, and $dE_b/dZ=s$, where $s$ in given in Eq. (\ref{eq4}). The coefficients $a_0$ and $a_{-1}$ are determined from these conditions.

\begin{figure}[t]
\begin{center}
\includegraphics[width=.95\linewidth,angle=0]{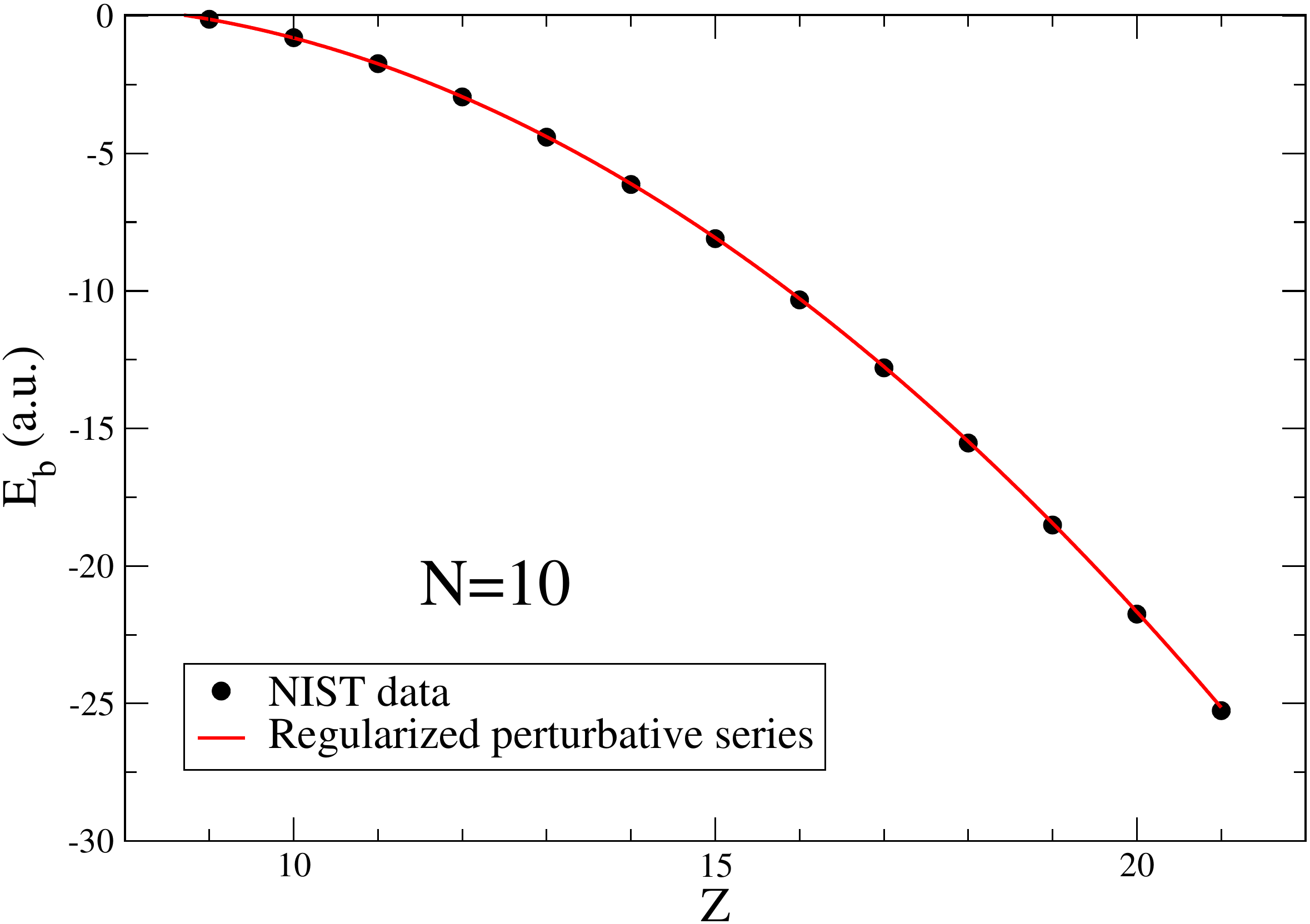}
\caption{\label{fig6} (Color online) The ionization potentials of Ne-like ions \cite{NIST} and the regularized perturbative series given by Eq. (\ref{eq10}).}
\end{center}
\end{figure}

In Fig. \ref{fig6}, we draw the renormalized perturbative series for $N=10$, that is Ne-like ions, and compare with the experimental data \cite{NIST}. In this case, the large-$Z$ expansion
leads to $a_2=1/8$, $a_1=1.6365$, whereas in order to fix $a_0$ and $a_{-1}$, we use $E_a(F)=0.125$, $R_{cov}(F)=1.4173$, which give $s=-0.5472$. The results are surprisingly good. 

Notice that, if we compute $g_c$ from the regularized series, we obtain $g_c=0.242$, a number very close to Kais estimation, $g_c=0.248$.

Computations for other ions are to be presented elsewhere \cite{largeZ}.

\section{Conclusions}

We have shown that the properties of neutral atoms can not always be qualitatively understood in terms of the prescriptions of TF theory, which are, however, valid in the cationic domain, $Z>N$.

The anionic domain, $Z<N$, is characterized by an instability threshold at $Z=Z_c\lesssim N-1$, and a linear dependence of the atom binding energy on $Z$. The ionization potential of the neutral atom, for example, is mostly determined by this linear dependence. Energy gaps in other sectors of the spectrum, atomic susceptibilities, some aspects of chemical binding, etc could be qualitatively related to the anionic regime also.

We show that the slope of the energy curve can be computed at $Z=N-1$ by a kind of zero-range forces theory. It is amazing that such a simple model provides quite accurate results for the slope in many atoms.

From the slope and the electron affinity at $Z=N-1$, we estimate both $Z_c$ and $I_p$ for a set of atoms. The estimations can be notably improved if we add information about the large-$Z$ regime. Indeed, we write a regularized perturbative series for the binding energy which reproduces not only $I_p$, but the ionization potential of cations as well. The large-$Z$ asymptotic behavior is analytically, and very simply, computed. Relativistic corrections could be straightforwardly incorporated in order to consider highly ionized cations. Research along these directions is currently in progress. 

\begin{acknowledgments}
The authors are grateful to the participants of the Theory Seminar at
the ICIMAF for discussions and remarks. Support by the Caribbean 
Network for Quantum Mechanics, Particles and Fields (ICTP, Trieste, 
Italy) is acknowledged.
\end{acknowledgments}

\appendix

\section{The interpolation scheme by Kais and co-workers}

A very rough estimation of $Z_c$ would consist of a linear interpolation up to $E_b=0$ from
the points $E_b=-I_p$ at $Z=N$, and $E_b=-E_a$ at $Z=N-1$.

Kais and co-workers \cite{Kais} suggest a more refined scheme, based on a Schrodinger
equation for the external orbital, which is similar to our Eq. (\ref{eq2}). They use an 
effective potential of the form:

\begin{equation}
V_{eff}=V_K(r)+\frac{\alpha}{r},
\end{equation}

\noindent
where $\alpha$ has the same meaning as in Eq. (\ref{eq2}), and the short range part is given by:

\begin{equation}
V_K(r)=-\frac{N-1}{r} e^{-\delta r}.
\end{equation}

\noindent
Notice that the Coulomb component of $V_{eff}$ has the correct behavior at $r\to 0$ and
$r\to\infty$. The short-range component, on the other hand, is controlled by the 
parameter $\delta$. If we require that $E_b=-I_p$ at $Z=N$, we get a value $\delta_0$
for $\delta$. In the same way, by requiring that $E_b=-E_a$ at $Z=N-1$, we get a value
$\delta_1$. For $Z<N-1$, Kais and co-workers use a linear interpolation:

\begin{equation}
\delta(Z)=(Z-N+1)\delta_0+(N-Z)\delta_1.
\end{equation}

The use of an effective Schrodinger equation for the external orbital is justified at
$Z\approx N-1$, but controversial at $Z=N$. Nevertheless, by using this scheme, the
authors reproduce Configuration Interacion calculations by Hogreve \cite{Hogreve} 
for light atoms, $N<19$.

Still more controversial, in certain cases, is the definition of the external orbital
in the interval $Z_c<Z<N$. For example, Cr ($Z=N=24$) ionizes by losing a $4s$ electron. 
But, at $Z_c$ it loses a $3d$ electron (a second $3d$ electron moves to a $4s$ orbital
at the threshold).

We give in Table \ref{tab1} results for $g_c$, along with the used parameters, for the
atoms studied in Ref. \onlinecite{Kais}. Notice that for some atoms, indicated by  
an asterisk, $g_c$ was recomputed using our data for $I_p$ and $E_a$. In the particular case of 
Cu, the Kais scheme does not lead to a solution for $g_c$, that is the $E_b$ vs $Z$ curve
does not intersect the axis. Thus, a linear interpolation is employed in order to find $g_c$.

\begin{widetext}

\begin{table}
\begin{tabular}{|c|c|c|c|c|c|c|c|c|}
\hline
N & neutral atom & nl & $I_p(N)$ & $E_a(N-1)$ & $R_{cov}(N-1)$ & $\delta_0/N$ & $\delta_1/(N-1)$ & $g_c$  \\
\hline
2 & He * & 1s & 0.903568 & 0.027706 & .604712 & 1.066567 & 0.8809 & 0.087538 \\
\hline
4 & Be * & 2s & 0.342602 & 0.022705 & 2.456644 & 0.338749 & 0.258051 & 0.135131 \\
\hline
6 & C & 2p & 0.413808 & 0.010276 & 1.58737 & 0.255 & 0.218 & 0.039 \\
\hline
9 & F & 2p & 0.640276 & 0.053676 & 1.417295 & 0.239 & 0.215 & 0.124 \\
\hline
10 & Ne & 2p & 0.792481 & 0.124985 & 1.133836 & 0.232 & 0.211 & 0.248 \\
\hline
12 & Mg & 3s & 0.280993 & 0.020129 & 3.023562 & 0.162 & 0.130 & 0.12 \\
\hline
14 & Si & 3p & 0.299568 & 0.015901 & 2.34326 & 0.128 & 0.112 & 0.075 \\
\hline
15 & P * & 3p & 0.385378 & 0.051082 & 2.154288 & 0.122911 & 0.109656 & 0.206333 \\
\hline
16 & S & 3p & 0.380723 & 0.027452 & 2.059801 & 0.124 & 0.111 & 0.1 \\
\hline
17 & Cl & 3p & 0.47655 & 0.076328 & 1.965315 & 0.120 & 0.109 & 0.242 \\
\hline
18 & Ar * & 3p & 0.579154 & 0.132775 & 1.889726 & 0.116817 & 0.107827 & 0.370993 \\
\hline
20 & Ca & 4s & 0.224654 & 0.018422 & 3.779452 & 0.0897 & 0.0748 & 0.133 \\
\hline
21 & Sc * & 3d & 0.24113 & 0.000902 & 3.288123 & 0.0994536 & 0.091335 & 0.006485 \\
\hline
22 & Ti & 4p & 0.250644 & 0.006906 & 3.004665 & 0.0764 & 0.0675 & 0.042 \\
\hline
23 & V * & 4s & 0.247682 & 0.002902 & 2.721206 & 0.0887704 & 0.077378 & 0.01875 \\
\hline
24 & Cr * & 4s & 0.248664 & 0.019287 & 2.721206 & 0.0893301 & 0.075405 & 0.123894 \\
\hline
25 & Mn * & 4s & 0.273195 & 0.024467 & 2.456644 & 0.0871283 & 0.0751374 & 0.135467 \\
\hline
27 & Co * & 4s & 0.289008 & 0.005547 & 2.34326 & 0.08669 & 0.077199 & 0.028083 \\
\hline
28 & Ni * & 4s & 0.28067 & 0.02432 & 2.229877 & 0.0879596 & 0.0755515 & 0.136781 \\
\hline
29 & Cu * & - & 0.284091 & 0.042467 & 2.21098 & 0.0880935 & 0.0746685 & 0.190977 \\
\hline
30 & Zn & 4s & 0.34523 & 0.045369 & 2.305466 & 0.0839 & 0.0748 & 0.183 \\
\hline
32 & Ge & 4p & 0.290298 & 0.015797 & 2.324363 & 0.0745 & 0.0676 & 0.054 \\
\hline
33 & As & 4p & 0.360702 & 0.045312 & 2.267671 & 0.0727 & 0.0670 & 0.186 \\
\hline
34 & Se & 4p & 0.358393 & 0.029546 & 2.267671 & 0.0728 & 0.0673 & 0.113 \\
\hline
35 & Br & 4p & 0.434149 & 0.07427 & 2.229877 & 0.0715 & 0.0667 & 0.253 \\
\hline
36 & Kr & 4p & 0.514475 & 0.123625 & 2.21098 & 0.0704 & 0.0661 & 0.386 \\
\hline
38 & Sr * & 5s & 0.209280 & 0.017851 & 4.062911 & 0.0572688 & 0.0489107 & 0.161547 \\
\hline
39 & Y * & 4d & 0.22848 & 0.001763 & 3.59048 & 0.0618596 & 0.0580956 & 0.01255 \\
\hline
40 & Zr & 5p & 0.243789 & 0.011278 & 3.325918 & 0.0487 & 0.0450 & 0.058 \\
\hline
41 & Nb & 4d & 0.248381 & 0.01565 & 3.099151 & 0.0614 & 0.0580 & 0.088 \\
\hline
42 & Mo * & 5s & 0.260642 & 0.03366 & 2.947973 & 0.0546169 & 0.0485206 & 0.216432 \\
\hline
43 & Tc * & 5s & 0.267533 & 0.027479 & 2.759 & 0.054444 & 0.04876 & 0.149862 \\
\hline
44 & Ru * & 5s & 0.270491 & 0.020205 & 2.607822 & 0.0544527 & 0.0484811 & 0.245631 \\
\hline
45 & Rh * & 5s & 0.274107 & 0.038573 & 2.570028 & 0.054438 & 0.0485292 & 0.241923 \\
\hline
46 & Pd * & 4d & 0.306373 & 0.041769 & 2.532233 & 0.0610524 & 0.0579478 & 0.210075 \\
\hline
47 & Ag * & 5s & 0.278419 & 0.020646 & 2.456644 & 0.0545197 & 0.0491283 & 0.092332 \\
\hline
48 & Cd * & 5s & 0.330514 & 0.047831 & 2.570028 & 0.0528306 & 0.048442 & 0.209007 \\
\hline
50 & Sn & 5p & 0.269882 & 0.011021 & 2.683411 & 0.0486 & 0.0450 & 0.055 \\
\hline
51 & Sb * & 5p & 0.31635 & 0.040853 & 2.645617 & 0.0479284 & 0.0447179 & 0.187077 \\
\hline
52 & Te & 5p & 0.31635 & 0.038426 & 2.645617 & 0.0478 & 0.0447 & 0.167 \\
\hline
53 & I * & 5p & 0.384074 & 0.072433 & 2.588925 & 0.0471871 & 0.0444753 & 0.282776 \\
\hline
54 & Xe * & 5p & 0.445763 & 0.112416 & 2.570028 & 0.0466429 & 0.0441987 & 0.402355 \\
\hline
57 & La & 6p & 0.204948 & 0.005313 & 3.892836 & 0.0349 & 0.0321 & 0.046 \\
\hline
58 & Ce & 5d & 0.20354 & 0.017266 & 3.666069 & 0.0419 & 0.0400 & 0.095 \\
\hline
60 & Nd * & 4f & 0.203039 & 0.004042 & 3.212534 & 0.0513798 & 0.0498267 & 0.03126\\
\hline
70 & Yb * & 6s & 0.229826 & 0.009334 & 3.344815 & 0.0384843 & 0.0345 & 0.06 \\
\hline
74 & W * & 6s & 0.288997 & 0.011829 & 2.985767 & 0.0370376 & 0.0342212 & 0.15 \\
\hline
75 & Re & 5d & 0.287874 & 0.029987 & 2.834589 & 0.0415 & 0.0400 & 0.116 \\
\hline
78 & Pt & 5d & 0.329227 & 0.042482 & 2.494438 & 0.0412 & 0.0399 & 0.178 \\
\hline
79 & Au * & 5d & 0.339029 & 0.078203 & 2.456644 & 0.041449 & 0.03989 & 0.421454 \\
\hline
80 & Hg * & 6s & 0.383570 & 0.084854 & 2.456644 & 0.0358645 & 0.0337502 & 0.349759 \\
\hline
82 & Pb & 6p & 0.272554 & 0.013855 & 2.721206 & 0.0340 & 0.0321 & 0.054 \\
\hline
83 & Bi & 6p & 0.267735 & 0.013372 & 2.740103 & 0.0341 & 0.0321 & 0.071 \\
\hline
84 & Po & 6p & 0.309206 & 0.034619 & 2.834589 & 0.0338 & 0.0320 & 0.163 \\
\hline
86 & Rn * & 6p & 0.394997 & 0.102898 & 2.796795 & 0.0332561 & 0.031709 & 0.428112 \\
\hline
89 & Ac & 7p & 0.189993 & 0.003675 & 3.987322 & 0.0256 & 0.0240 & 0.042 \\
\hline
\end{tabular}
\caption{Data used in the present work for $I_p$, $E_a$ and $R_{cov}$, along with the quantum numbers, $nl$, of the external orbital, the
parameters $\delta_0$, $\delta_1$, and the estimated values of $g_c$ for the atoms considered by Kais et al \cite{Kais}.
An asterisk near the atom means that we recomputed $g_c$ using our data for $I_p$ and $E_a$. In the particular case of Cu, no solution 
is found, thus a linear interpolation is employed.}
\label{tab1}
\end{table}

\end{widetext}

\end{document}